\documentclass[twocolumn,showpacs,preprintnumbers,amsmath,amssymb]{revtex4}
\usepackage[dvips]{graphicx} 
\usepackage[english]{babel}
\usepackage{amsfonts}
 \usepackage{amssymb}

\begin{document}

\title{Interacting universes and the cosmological constant}

\author{A. Alonso-Serrano$^{1,2}$, C. Bastos$^{3}$, O. Bertolami$^{3,4}$,  and S. Robles-P\'{e}rez $^{1,2} $}

\affiliation{$^1$ Centro de F\'{\i}sica ``Miguel Catal\'{a}n'', Instituto de  F\'{\i}sica Fundamental, Consejo Superior de Investigaciones Cient\'{\i}ficas, Serrano 121, 28006
Madrid, Spain \\
 $^2$ Estaci\'{o}n Ecol\'{o}gica de Biocosmolog\'{\i}a, Pedro de Alvarado, 14, 06411 Medell\'{\i}n, Spain\\ 
$^3$ Instituto de Plasmas e Fus\~ao Nuclear, Instituto Superior T\'ecnico, Avenida Rovisco Pais 1, 1049-001 Lisboa, Portugal \\   
  $^4$ Departamento de F\'isica e Astronomia, Faculdade de Ci\^ encias da Universidade do Porto, Rua do Campo Alegre 687, 4169-007 Porto, Portugal }

\date{\today}

\begin{abstract}
\end{abstract}

\maketitle

\noindent
{\bf Dedicated to the memory of our friend and colleague Pedro Gonz\'alez-D\'\i az.}

\section{Introduction}

The idea of a multitude of universes has many avatars in modern physics. It has first appeared in the context of the many-worlds interpretation of quantum mechanics \cite{Everett1957} and later when it was realized that inflation could lead, due to quantum effects in the early universe, to the creation new universes  \cite{Linde1986}. More recently, the concept reemerged in the context of string theory when it was understood that the initial outlook about the original distinct string theories was not correct and these should be seen as a continuum of theories. This interpretation suggests that one has actually different solutions of a more fundamental theory. The space of these solutions is referred to as supermoduli-space. The moduli are fields, and their variation allows 
moving in the supermoduli-space. The moduli vary as one moves in the spacetime, as they have their own equations of motion.
The continuum of solutions in the supermoduli-space are supersymmetric and thus, have all a vanishing cosmological constant. 
It follows that in order to describe our world, there must exist some non-supersymmetric ``islands'' in the supermoduli-space. 
It is believed that the number of these discrete vacua is huge, $G = 10^{100}$, or googleplexes $10^G$ \cite{Susskind1}. Since the magnitude of the 
cosmological constant is about $10^{120}$ smaller than its natural value $M_P^4$, where $M_P= 1.2 \times 10^{19}$ GeV 
is the Planck mass, it is highly unlikely to find such a vacuum, unless there exists a huge number of solutions with 
every possible value for the cosmological constant. 
The space of all such string theory vacua is called the {\it landscape}  
\cite{Bousso2000} and suggests a {\it multiverse}. According to this proposal, the multiple vacua of string theory is associated 
to a vast number of ``pocket 
universes'' in a single large Mega-universe \cite{Susskind1}. Thus, the vacuum associated to our expanding universe must emerge from a selection procedure, through anthropic reasoning \cite{PolL} or quantum cosmological arguments (see e.g. \cite{Mersini2008}). 
For sure, this interpretation is somewhat disturbing and not free from criticism as the impossibility of observing a multiverse implies that its scientific status is questionable, unless it leads to unequivocal phenomena that cannot be accounted in the context of the usual single universe framework \cite{GEllis06}.  It is interesting to point out that recently, it has been suggested that the many-worlds interpretation of quantum mechanics could be regarded as a quantum mechanics of the multiverse \cite{BoussoSusskind2011,Nomura2012} (see however, \cite{OBHerdeiro2012}). 

Naturally, the multiverse proposal poses many intriguing questions. For instance, do the universes interact with each other? Do they mutually affect their properties? Do they exhibit collective behavior features? The aim of this work is precisely to address these issues.  

\section{Interaction scheme in a third quantization formalism}

In Ref. \cite{Bertolami2008}, it has been proposed an interaction scheme between universes which follows from the so-called Curvature Principle. The main ingredients of the interaction scheme through the Curvature Principle are the following: i) the state of each single universe in the multiverse is described by a curvature invariant, $I$; ii) the vacuum energy of each single universe depends on the matter fields of the universe as well as on the interaction with other universes; and, iii) it can be posed a 'meta cosmic time' to describe the evolution of the curvature invariants.

As a result of the interaction between two universes, it is shown that the value of the cosmological constant of one of the universes effectively vanishes at the expense of an increasing value of the cosmological constant of the partner  \cite{Bertolami2008}. The multiverse thus opens the door to novel ways of facing up the problem of the cosmological constant (see also \cite{Linde1988, RP2011}).

In the third quantization formalism of quantum cosmology \cite{McGuigan1988, Rubakov1988, Strominger1990}, the description of the wave function of the universe is naturally extended to a many-universe scenario much in the same way as a many-particle description of fields arises in a quantum field theory. In particular, for homogeneous and isotropic universes, the Wheeler-De Witt equation can formally be seen as a wave equation defined upon the minisuperspace spanned by the scale factor, $a$, and the matter fields, $\vec{\varphi}=(\varphi_1, \ldots, \varphi_n)$. The third quantization formalism consists in considering the wave function of the universe as the field to be quantized that propagates on minisuperspace and, thus,  the quantum state of the universe can be studied as a quantum field theory in the minisuperspace.

We point out that assuming homogeneity and isotropy are essential conditions as far as it concerns large macroscopic universes, for which the quantum fluctuations of the metric are negligible with respect to the value of the metric. Then, in a first approach, an homogeneous and isotropic universe can properly model our universe for most of its history.

Thus, let us consider a closed homogeneous and isotropic universe endowed with a slow-varying field, $\varphi \approx \varphi_0$. The Wheeler-De Witt equation, with an appropriate choice of factor ordering, can then be written as
\begin{equation}\label{WDW1}
\hbar^2 \frac{\partial^2\phi}{\partial a} + \frac{\hbar^2}{a} \frac{\partial \phi}{\partial a} + (a^4 V(\varphi_0) - a^2) \phi = 0 , 
\end{equation}
where $\phi\equiv \phi(a,\varphi)$ is the wave function of the universe, and $V(\varphi_0)$ is the potential of the scalar field evaluated at $\varphi_0$. The wave function $\phi(a,\varphi_0)$ effectively represents the quantum state of a de Sitter universe with a value of the cosmological constant given by, $\Lambda \equiv V(\varphi_0)$. Furthermore, Eq. (\ref{WDW1}) can formally be seen as the equation of a harmonic oscillator with a variable friction term and variable frequency. Indeed, this can be made clearer rewriting:
\begin{equation}\label{WDW2}
\ddot{\phi} + \frac{\dot{M}}{M} \dot{\phi} + \omega^2 \phi = 0 , 
\end{equation}
where, $\dot{\phi}\equiv \frac{\partial \phi}{\partial a}$, $M\equiv M(a) = a$, and $\omega\equiv \omega(a) = \frac{a}{\hbar} \sqrt{a^2 \Lambda - 1}$. In the third quantization formalism, the scale factor formally plays the role of an intrinsic time variable of the minisuperspace and, then, the quantum state of the universe can be described in the basis of number eigenstates of the harmonic oscillator with \emph{time} dependent friction and frequency, $M(a)$ and $\omega(a)$, respectively. The Hamiltonian for which the Heisenberg equations of motion give rise to the Wheeler-de Witt equation (\ref{WDW2})  reads
\begin{equation}\label{Hamiltonian}
H = \frac{1}{2 M} p^2_\phi +\frac{M \omega^2}{2} \phi^2 . 
\end{equation}

An interaction scheme in the multiverse can now be easily posed by following the analogy with quantum mechanics. For instance, let us consider the interaction between two universes described by a total Hamiltonian, $H_T$, given by
\begin{equation}\label{HamiltonianT}
H_T = H_1 + H_2 + H_I ,
\end{equation}
where $H_1$ and $H_2$ are given by Eq. (\ref{Hamiltonian}), with $(\omega_1, \phi_1, p_{\varphi_1})$ and $(\omega_2, \phi_2, p_{\varphi_2})$ for the universes labelled by indices $1$ and $2$, respectively. The \emph{mass}, $M(a)$, turns out to be the same for both universes because it arises from a particular choice of the factor ordering, which is assumed to be identically chosen in both cases. 

For a large parent universe the geometric term in Eq. (\ref{WDW1}) can be disregarded and, thus, $\omega_1^2 \approx a^4 \Lambda_1$ and $\omega_2^2 \approx a^4 \Lambda_2$. The frequency of the harmonic oscillator turns out then to be  proportional to the vacuum energy of each single universe. Furthermore, let us assume an interaction Hamiltonian, $H_I$, given by
\begin{equation}
H_I = \frac{M a^4 k}{2} (\phi_2 - \phi_1 )^2 ,
\end{equation}
where $k$ is a coupling constant to be subsequently determined. Following the procedure used in Refs. \cite{Macedo2012, Menouar2010}, we can apply the canonical transformation derived from the following generating function,
\begin{widetext}
\begin{equation}
G(\phi_1, \phi_2, P_1, P_2, a) = \phi_1 (P_1 \cos\theta + P_2 \sin\theta) + \phi_2 (- P_1 \sin\theta + P_2 \cos\theta) .
\end{equation}
\end{widetext}
Then, the new canonical variables, $(\Phi_1, P_1)$ and $(\Phi_2, P_2)$, can be obtained through the following equations,
\begin{eqnarray}\label{tran1}
p_{\phi_1} &\equiv& \frac{\partial G}{\phi_1} = P_1 \cos\theta + P_2 \sin\theta , \\
p_{\phi_2} &\equiv& \frac{\partial G}{\phi_2} = - P_1 \sin\theta + P_2 \cos\theta , \\
\Phi_1 &\equiv& \frac{\partial G}{P_1} = \phi_1 \cos\theta - \phi_2 \sin\theta , \\ \label{tran2}
\Phi_2 &\equiv& \frac{\partial G}{P_2} = \phi_1 \sin\theta + \phi_2 \cos\theta .
\end{eqnarray}
The Hamiltonian (\ref{Hamiltonian}) is transformed according to $H \rightarrow H_N = H + \frac{\partial G}{\partial a}$, with
\begin{widetext}
\begin{equation}
H_N = \frac{1}{2 M} P_1^2 + \frac{M \Omega_1^2}{2} \Phi_1^2 +\frac{1}{2 M} P_2^2 + \frac{M \Omega_2^2}{2} \Phi_2^2 + \Phi_1\Phi_2 \left( (b_1-b_2) \sin2\theta + b_3 \cos2\theta \right) + \dot{\theta} (\Phi_1 P_2 - P_1 \Phi_2) ,
\end{equation}
\end{widetext}
where
\begin{eqnarray}
b_1 \equiv b_1(a) &=& M a^4 ( \Lambda_1 + k ) , \\
b_2 \equiv b_2(a) &=& M a^4  (\Lambda_2 + k ) , \\
b_3 \equiv b_3(a) &=& - 2 M a^4 k ,
\end{eqnarray}
and
\begin{eqnarray}\nonumber
\Omega_1^2 &=& \frac{1}{M} (b_1 \cos^2\theta + b_2 \sin^2\theta - \frac{b_3}{2} \sin2\theta) , \\ \label{Omega1}
 &=& \frac{1}{2 M} (  (b_1-b_2) \cos2\theta - b_3 \sin2\theta + b_1 +b_2) , \\ \nonumber
\Omega_2^2 &=& \frac{1}{M} (b_1 \sin^2\theta + b_2 \cos^2\theta + \frac{b_3}{2} \sin2\theta) ,\\  \label{Omega2}
 &=& \frac{1}{2 M} (  (b_2-b_1) \cos2\theta + b_3 \sin2\theta + b_1 +b_2) .
\end{eqnarray}
If the canonical transformation is such that, $\dot{\theta}\approx 0$, and
\begin{equation}
\tan2\theta = \frac{b_3}{b_2 - b_1} (\approx {\rm const.}) ,
\end{equation}
then, the Hamiltonian $H_N$ represents the dynamical evolution of two non-interacting universes, as they would be seen from internal observers, for which the vacuum energy is now associated to the new frequencies $\Omega_1$ and $\Omega_2$, respectively. By choosing an appropriate coupling constant, $k$, we can obtain a pair of universes for which the vacuum energy of one of them would approximately be zero at the expense of an increasing value of the vacuum energy of the partner universe, the behaviour obtained in Ref. \cite{Bertolami2008} through the proposed Curvature Principle.

Indeed, following Ref. \cite{Bertolami2008}, let us consider two ``nearby" universes, i.e. two interacting universes with similar values of their cosmological constant, given by
\begin{eqnarray}\label{l1}
\Lambda_1 &=& \Lambda + \varepsilon , \\ \label{l2}
\Lambda_2 &=& \Lambda - \varepsilon ,
\end{eqnarray}
where, $\varepsilon\ll 1$, may represent some small fluctuation of the vacuum energy in each single universe. Then, 
\begin{eqnarray}
b_1 - b_2 &=& 2 M a^4 \varepsilon , \\
b_1 + b_2 &=& 2 M a^4 (\Lambda + k) , \\
b_3 &=& - 2 M a^4 k,
\end{eqnarray}
and, disregarding orders higher than $\varepsilon$,
\begin{eqnarray}
\tan2\theta &=& -\frac{k}{\varepsilon} , \\
\cos2\theta &=& \frac{\varepsilon}{\sqrt{k^2 + \varepsilon^2}} \approx \frac{\varepsilon}{| k|} , \\
\sin2\theta &\approx& 1 ,
\end{eqnarray}
with, $k<0$. Inserting these values into Eqs. (\ref{Omega1})-(\ref{Omega2}), it follows that
\begin{eqnarray}
\Omega_1^2 &\approx&  a^4 (\Lambda + 2 k + \varepsilon \cos2\theta ) \equiv a^4 \Lambda_1^{ef}  , \\
\Omega_2^2 &\approx&  a^4 (\Lambda - \varepsilon \cos2\theta ) \equiv a^4  \Lambda_2^{ef} .
\end{eqnarray}
Now, it can easily be checked that for a coupling constant, $k = - \frac{\Lambda}{2}$, the effective values of the vacuum energy of the universes read
\begin{eqnarray}
\Lambda_1^{ef} &\approx& \frac{2 \varepsilon^2}{\Lambda} \approx 0 , \\
\Lambda_2^{ef} &\approx& \Lambda - \frac{2 \varepsilon^2}{\Lambda} \approx \Lambda .
\end{eqnarray}
Therefore, the two interacting universes would be seen, in the new representation, as two non-interacting universes with an effective value of their cosmological constants given by $\Lambda_1^{ef}$ and $\Lambda_2^{ef}$, respectively. Let us notice that a 'super-observer' that could make measurements within the whole multiverse would see the two universes as interacting to each other. However, an observer inside a universe that would consider her universe as an isolated non-interacting universe would then assume that the vacuum energy of her universe is given by $\Lambda_1^{ef}$ or $\Lambda_2^{ef}$, respectively.

\subsection{General quadratic potential}

The same procedure can also be applied to a general quadratic potential. Let us consider the following Hamiltonian
\begin{equation}
H = \frac{1}{2 M} (p_1^2 + p_2^2) + A \phi_1^2 + B \phi_2^2 + C \phi_1 \phi_2 ,
\end{equation}
which, as a particular case, includes the Hamiltonian given by Eq. (\ref{HamiltonianT}). Let us assume that, $A = \alpha a^4 M(a)$, $B=\beta a^4 M(a)$, and $C=\gamma a^4 M(a)$, where $\alpha$, $\beta$, and $\gamma$ are constants coefficients and, $\alpha \propto \Lambda_1$ and $\beta\propto \Lambda_2$, with $\Lambda_1$ and $\Lambda_2$ being the value of the cosmological constants of the universes $1$ and $2$, respectively, and $\gamma$ being the coupling constant of the interaction between the universes.

The kinetic term is invariant under the canonical transformation given by Eqs. (\ref{tran1}-\ref{tran2}), i.e. $p_1^2 + p_2^2 = P_1^2 + P_2^2$, however, the potential term transforms into
\begin{equation}
\frac{M\Omega_1^2}{2} \Phi_1^2 + \frac{M\Omega_2^2}{2} \Phi_2^2 + \left( \sin2\theta (A-B)+ C \cos2\theta \right) \Phi_1 \Phi_2 , 
\end{equation}
where, $\Omega_1^2 = \tilde{\Lambda}_1 a^4$ and $\Omega_2^2 = \tilde{\Lambda}_2 a^4$, with
\begin{eqnarray}\label{L1}
\tilde{\Lambda}_1 &=& (\alpha - \beta) \cos2\theta - \gamma \sin2\theta + (\alpha + \beta) , \\ \label{L2}
\tilde{\Lambda}_2 &=& (\beta-\alpha) \cos2\theta + \gamma \sin2\theta + (\alpha + \beta ) .
\end{eqnarray}
Therefore, considering once again a constant value $\theta$ so that
\begin{equation}\label{L3}
\tan2\theta = \frac{\gamma}{\beta - \alpha} ,
\end{equation}
then, $\dot{\theta} = 0$ and the transformed Hamiltonian turns out to be
\begin{equation}
H_N = \frac{1}{2 M} P_1^2 + \frac{M \Omega_1^2}{2} \Phi_1^2 +\frac{1}{2 M} P_2^2 + \frac{M \Omega_2^2}{2} \Phi_2^2 .
\end{equation}
This represents the dynamics, in the transformed variables, of two non-interacting universes with effective values of their cosmological constants given by $\tilde{\Lambda}_1$ and $\tilde{\Lambda}_2$, respectively.

We can now analyze the value of the coupling constant that makes one of the effective cosmological constants vanishingly small, let us say $\tilde{\Lambda}_1=0$. Then, we search for the value of $\gamma$ that solves the system of equations (\ref{L1}-\ref{L3}). Summing up the first two equations, with $\tilde{\Lambda}_1=0$, we obtain
\begin{equation}
\tilde{\Lambda}_2 = 2 (\alpha + \beta) .
\end{equation}
From Eqs. (\ref{L2})-(\ref{L3}):
\begin{equation}
\frac{(\beta-\alpha)^2 + \gamma^2}{\beta-\alpha} \cos2\theta = \frac{\tilde{\Lambda}_2}{2} .
\end{equation}
From Eq. (\ref{L3}), follows that, $\cos2\theta = \frac{\beta-\alpha}{\sqrt{\gamma^2 + (\beta - \alpha)^2}}$, so we finally obtain 
\begin{equation}
\gamma^2 = \frac{\tilde{\Lambda}_2^2}{4} - (\beta -\alpha )^2 = 4 \alpha \beta .
\end{equation}
Therefore, for a value of the coupling constant $\gamma^2 = 4 \alpha \beta$, the interaction between the universes yields, in the new representation, the values $\tilde{\Lambda}_1 = 0$ and $\tilde{\Lambda}_2 = 2(\alpha +\beta)$, for the two non-interacting universes. As a particular case, we have the values $\alpha = \frac{\Lambda_1 + k}{2}$, $\beta=\frac{\Lambda_2 + k}{2}$, and $\gamma = -k$, of the preceding section, with $\Lambda_1$ and $\Lambda_2$ given by Eqs. (\ref{l1}) and (\ref{l2}), respectively, and $k=-\frac{\Lambda}{2}$.

\section{A multiverse of interacting harmonic oscillators}

In this section, it is shown that, as it occurs in other branches of physics, the multiverse as a whole might exhibit collective phenomena that make it being physically richer than the sum of its parts. This sheds a new perspective into the customary problems in cosmology and, particularly, provides a new perspective to the problem of the cosmological constant.

Let us consider a multiverse of $N$ interacting de-Sitter universes represented, in the third quantization formalism, by harmonic oscillators like those described in the preceding section, with scale factor dependent mass and frequency, given respectively by $\mathcal{M}(a) = a$ and $\omega^2(a)\approx \Lambda a^4$, where $\Lambda$ is the value of cosmological constant of the de-Sitter universes. Following references \cite{Lievens2006, Lievens2008a, Lievens2008b}, let us assume in the multiverse some kind of 'nearest interaction' described by a total Hamiltonian given by
\begin{equation}\label{Hamiltonian31}
\hat{H} = \sum_{r=1}^N \left( \frac{\hat{p}_r^2}{2 \mathcal{M}} + \frac{\mathcal{M} \omega^2}{2} \hat{\phi}_r^2 + \frac{\mathcal{M} c}{2} (\hat{\phi}_r - \hat{\phi}_{r+1}) ^2 \right) ,
\end{equation}
where $\hat{\phi}_r$ is the wave function operator of the $r$-universe. There exists then a finite Fourier transform, given by \cite{Lievens2006, Plenio2004}
\begin{eqnarray}\label{transf1}
\hat{\Phi}_r &=& \frac{1}{\sqrt{N}} \sum_{k=1}^N e^{-(2 \pi i r k/N)} \hat{\phi}_k , \\ \label{transf2}
\hat{P}_r &=& \frac{1}{\sqrt{N}} \sum_{k=1}^N e^{(2 \pi i r k/N)} \hat{p}_k ,
\end{eqnarray}
for which the Hamiltonian (\ref{Hamiltonian31}) transforms into
\begin{equation}
\hat{H} = \sum_{r=1}^N \left( \frac{1}{2 \mathcal{M}} \hat{P}_r \hat{P}_r^\dag + \frac{\mathcal{M} \omega^2_r}{2} \hat{\Phi}_r \hat{\Phi}_r^\dag \right) , 
\end{equation}
where the frequency of the \emph{normal modes}, $\omega_r$, is given by
\begin{equation}\label{nmodes}
\omega_r^2 = \omega^2 + 4 c \sin^2\left( \frac{\pi r}{N}\right) ,
\end{equation}
with $\omega_{N-r} = \omega_r$ and $c$ a constant. In the transformation given by Eqs. (\ref{transf1}-\ref{transf2}) it has been imposed a periodic boundary condition such that $\hat{\phi}_{N+1} = \hat{\phi}_1$ which provides a ``closed form" for the multiverse. Notice that we could have imposed instead a ``fixed wall" boundary condition, see Ref. \cite{Lievens2008a}.

It is worth pointing out that it would be enough to take an appropriate negative value of the coupling constant $c$ to obtain a normal mode with a value of the frequency close to zero. However, in order to satisfy the canonical commutation relations, some algebraic conditions have to be satisfied and an appropriate representation has to be chosen. The energy eigenstates of the Hamiltonian in such representation would yield the energy levels of the normal modes. Following Refs. \cite{Lievens2006, Lievens2008a, Lievens2008b}, for a given representation, the energy spectrum splits into a large number of different levels, like in other collective phenomena (crystals, phonons,...). Given some conditions, the ground state of the new spectrum approaches to zero. Therefore, we could say that, as a quantum collective system, the multiverse would posses normal modes for which the ground state turns out to be greatest even though close to zero.

Let us define the following operators,
\begin{eqnarray}
a_r^- &\equiv& \sqrt{\frac{\mathcal{M} \omega_r}{2\hbar}} \hat{\Phi}_r + \frac{i}{\sqrt{2 \mathcal{M} \omega_r \hbar}} \hat{P}_r^\dag , \\
a_r^+ &\equiv& \sqrt{\frac{\mathcal{M} \omega_r}{2\hbar}} \hat{\Phi}_r^\dag - \frac{i}{\sqrt{2 \mathcal{M} \omega_r \hbar}} \hat{P}_r ,
\end{eqnarray}
with $(a^\pm_r)^\dag = a^\mp_r$. In terms of the operators $a^\pm_r$, the Hamiltonian takes the form
\begin{equation}\label{Hamiltonian32}
\hat{H} = \sum_{r=1}^N \frac{\hbar \omega_r}{2} \{a^-_r , a^+_r\} = \sum_{r=1}^N \frac{\hbar \omega_r}{2} (a^-_r a^+_r + a^+_r a^-_r ) .
\end{equation}
The condition the operators $a^\pm_r$ must satisfy in order to comply with the canonical commutation relations turns out to be equivalent than to require that \cite{Lievens2006, Lievens2008a, Lievens2008b}
\begin{equation}\label{condition}
[\hat{H}, a_r^\pm] = \pm \hbar \omega_r a_r^\pm .
\end{equation}
In the these Refs., it is shown the conditions that have to be satisfied for the class of appropriate representations for which the operators $a^\pm_r$ satisfy Eq. (\ref{condition}).

We can represent the ladder operators $a^\pm_r$ in terms of the basis elements of the Lie superalgebra $\mathfrak{gl}(1|N)$, $e_{jk}$ with $j,k=1, 2, \ldots, N$, as
\begin{equation}
a_r^- = \sqrt{\frac{2\beta_r}{\omega_r}} e_{r 0} \;\; , \; \; a_r^+ =  \sqrt{\frac{2\beta_r}{\omega_r}} e_{0 r} ,
\end{equation}
with $(e_{0k})^\dag = e_{k 0}$, following then that $(a^\pm)^\dag = a^\mp$. The elements $e_{r 0}$ and $e_{0r}$ satisfy the anticommutation relation (see, Eq. (3.1) of Ref. \cite{Lievens2008a})
\begin{equation}
\{e_{r0}, e_{0r} \} = e_{rr} + e_{00} ,
\end{equation}
and the Hamiltonian (\ref{Hamiltonian32}) can therefore be written as
\begin{equation}\label{Hamiltonian33}
\hat{H} = \hbar \left( \beta e_{00} + \sum_{k=1}^N \beta_k e_{kk} \right) ,
\end{equation}
with
\begin{equation}
\beta_k \equiv - \omega_k + \frac{1}{N-1} \sum_{j=1}^N \omega_j \;\; , \;\; \beta \equiv \sum_{j=1}^N \beta_j .
\end{equation}
Thus, Eq. (\ref{condition}) is satisfied and so are the canonical commutation relations. Furthermore, in terms of the so-called ladder representation \cite{Lievens2008a}, $V(p)$, characterized by a positive integer $p$, the orthogonal basis elements of $V(p)$ are described in terms of a fermionic variable $\theta$, with $\theta \in \{0,1\}$, and $N$ bosonic variables, $s_i$, with $i=1,2,\ldots, N$ and $s_i=\{0,1,2,\ldots\}$, given by
\begin{equation}
w(\theta;s)\equiv w(\theta;s_1, s_2, \ldots, s_N) ,
\end{equation}
with, $\theta + s_1 + s_2 +\ldots + s_N = p$. The action of the algebra elements $e_{00}$, $e_{kk}$, $e_{k0}$, and $e_{0k}$ over the basis elements $w(\theta;s)$ reads \cite{Lievens2008a}
\begin{eqnarray}
e_{00} w(\theta; s) &=& \theta w(\theta; s) , \\
e_{kk} w(\theta; s) &=& s_k  w(\theta; s) , \\
e_{k0} w(\theta; s) &=& \theta \sqrt{s_k + 1}  w(\theta; s_1, \ldots, s_k+1, \ldots, s_N ) , \\
e_{0k} w(\theta; s) &=& (1-\theta) \sqrt{s_k}  w(\theta; s_1, \ldots, s_k-1, \ldots, s_N ) .
\end{eqnarray}
Then, each basis vector $w(\theta; s)$ is an eigenvector of the Hamiltonian ($\ref{Hamiltonian33}$), i.e. $\hat{H} w(\theta;s) = \hbar E_{\theta,s} w(\theta; s)$, with eigenvalues $E_{\theta,s}$ given by
\begin{equation}
E_{\theta,s} = \beta \theta +\sum_{k=1}^M \beta_k s_k .
\end{equation}
These eigenvalues do provide us with a spectrum of the diagonalized Hamiltonian in terms of the normal modes of the set of universes, being these taken as a collective system.

\subsubsection{Energy spectrum: no interaction case.}

Let us now analyze the energy spectrum for different values of the coupling $c$. For $c= 0$, i.e. for no interaction among the universes, all the values $\omega_k$ in Eq. (\ref{nmodes}) are the same, $\omega_k = \omega$. The eigenvalues of the Hamiltonian then read
\begin{equation}
E_{\theta,s} = \omega \theta + \frac{\omega p}{N-1} .
\end{equation}
Therefore, there exist only two different eigenvalues, for $\theta = 0$ and $\theta=1$, given respectively by
\begin{eqnarray}
E_{0,s} &=& \frac{\omega p}{N-1} , \\
E_{1,s} &=& \frac{\omega p}{N-1}+\omega .
\end{eqnarray}
Let us notice that, for a fixed value of $p$, then $E_{0,s}\rightarrow 0$ and $E_{0,s}$, for a large number of universes in the multiverse, i.e. for $N\gg1$. Therefore, one of the new 'energy' levels turns out to be in that case close to zero, even though there is no interaction present among the universes. This is just a quantum effect having no classical analogue, that has come up because the consideration of the multiverse as a collective quantum system.

The multiplicities, $m_0$ and $m_1$, of the eigenvalues $E_{0,s}$ and $E_{1,s}$ are, respectively,
\begin{eqnarray}
m_0 &=& \binom{p+N-1}{N-1} , \\
m_1 &=& \binom{p+N-2}{N-1} . 
\end{eqnarray}
If the microcanonical states would be equally probable, then, the probabilities $P_{0,s}$ and $P_{1,s}$ of each eigenstate, with $\theta=0$ and $\theta=1$, respectively, would be proportional to the multiplicity of the state. Therefore, we would have
\begin{equation}
P_{0,s} = \frac{p}{p+N-1} P_{1,s} .
\end{equation}
It means that if $p<N-1$ and $N\gg 1$, the ground state of the multiverse would be much less probable that the excited state, i.e. $P_{0,s} \ll P_{1,s}$. Then, the collective vacuum  state, for which $E_{0,s}=0$, would be a very improbable state for the multiverse. However, if $p>N-1$ and $N\gg1$, then, $P_{0,s} = \varepsilon P_{1,s}$, with $\varepsilon \in (\frac{1}{2}, 1)$. In that case, the probability of a decay of the level $\theta=1$ into the ground state $\theta=0$, would be considerable and we might expect that the multiverse, may be initially created in the excited state, would finally decay into the ground state.

\subsubsection{Energy spectrum: the interacting case.}

For the case, $c_0 > c > 0$, where $c_0$ is a limiting value for the variables $\beta_k$ to be positive \cite{Lievens2006, Lievens2008a}, then, the two eigenvalues of the non-interacting case split into a number of different eigenvalues. Let us focus on the maximal and minimal eigenvalues. The former corresponds to the eigenstate $w(1;0,\ldots, p-1)$, with eigenvalue given by
\begin{equation}
E_{1,s} = \beta + (p-1) \beta_N .
\end{equation}
The assumed periodic boundary condition implies that, $\beta_k = \beta_{N-k}$, and therefore, $$\beta_1>\beta_2 > \ldots > \beta_{[\frac{N}{2}]} \leq \beta_{[\frac{N}{2}]+1} < \ldots < \beta_N ,$$and $\beta_1<\beta_N$, where $[\frac{N}{2}]$ is the integer part of $\frac{N}{2}$ and the equality between $\beta_{[\frac{N}{2}]}$ and $\beta_{[\frac{N}{2}]+1}$ only holds for $N$ even. Without lost of generality we can consider the latter case. Then, the minimal eigenvalue corresponds to the eigenvector $w(0;s_1,\ldots,s_N)$, with $s_i = p \, \delta_{i,\frac{N}{2}}$, for which
\begin{equation}
E_{0,s} = p \, \beta_\frac{N}{2} = p \left( \frac{1}{N-1} \sum_{j=1}^N \omega_j - \sqrt{\omega^2 + 4 c} \right) .
\end{equation}
Let us estimate an upper bound for the ground eigenvalue $E_{0,s}$. From the Cauchy-Schwartz inequality, for $c>0$,
\begin{equation}
\sum_{j=1}^N \omega_j \leq N^\frac{1}{2} \left( \sum_{j=1}^N \omega_j^2 \right)^\frac{1}{2} .
\end{equation}
On the other hand, with Eq. (\ref{nmodes}), 
\begin{equation}
\sum_{j=1}^N \omega_j^2 = N (\omega^2 + 2 c) .
\end{equation}
It then follows that
\begin{equation}
0 < \beta_\frac{N}{2} \leq \frac{N}{N-1} \sqrt{\omega^2 + 2 c} - \sqrt{\omega^2 + 4 c} \equiv \beta_{max}(c) ,
\end{equation}
where $\beta_{max}(c)$ is a function that vanishes for the value $c = c_0$, given by
\begin{equation}
c_0 = \frac{2 N - 1}{2(N^2 - 4N + 2)} \omega^2 .
\end{equation}
Therefore, for values $c < c_0$, it is obtained, $\beta_{max} > \beta_\frac{N}{2} > 0$. It implies that for a value of the interaction coupling,  $c \lesssim c_0$ , then, the energy of the ground state would be greater but fairly close to zero. 

For instance, let us take a value of the interaction coupling close to $c_0$, i.e.
\begin{equation}
c = \left( \frac{2 N - 1}{2(N^2 - 4N + 2)} - \frac{(N-1)^2}{N^2 - 4N + 2} \varepsilon \right) \omega^2 ,
\end{equation}
with $0 < \varepsilon < \frac{2N-1}{2(N-1)^2} \ll 1$ (the second inequality is needed to satisfy the condition $c>0$). Then, it is obtained that $\beta_{max}$ is of order $\varepsilon$, given by
\begin{equation}
\beta_{max} = \frac{\sqrt{N^2 - 4 N + 2}}{N} \, \omega \, \varepsilon ,
\end{equation}
so the ground state $E_{0,s}$ of the multiverse would be then of order $\varepsilon$, too.

Let us finally particularize to the more tractable case of three interacting universes (i.e. $N=3$). Then, $\omega_1 = \omega_2 = \sqrt{\omega^2 + 3 c}$ and $\omega_3 = \omega$. It yields the values
\begin{eqnarray}
\beta_1 = \beta_2 &=& \frac{\omega}{2} , \\
\beta_3 &=& \sqrt{\omega^2 + 3 c } -\frac{\omega}{2} , \\
\beta = \sum_{j=1}^3 \beta_j &=& \sqrt{\omega^2 + 3 c } +\frac{\omega}{2} .
\end{eqnarray}
For $N=3$ there is no restriction on the values of $c$ \cite{Lievens2006, Lievens2008a} as all the values of $\beta_j$ are positive, provided that $c>-\frac{\omega^2}{4}$. In that simple case, it can easily be seen that, for $c>0$, $\beta_3 > \beta_2 = \beta_1$, and the minimum value corresponds to $\beta_1 = \frac{\omega}{2}$. However, for values $0 > c > -\frac{\omega^2}{4}$, then $\beta_1 = \beta_2 > \beta_3$. Thus, the minimum eigenvalue is proportional to the value $\beta_3$, which can be fairly close to zero for $c\approx -\frac{\omega^2}{4}$. This fact might prompt us to check if there can also be a regime $c<0$ in the more general case for which $N >3$ (in the cited references, the authors restrict their attention to the general case $c>0$).

\section{Conclusions and Discussion}

The treatment of the multiverse as a quantum collective phenomena opens the route to novel approaches for traditional problems in cosmology. Besides providing a novel set of tools to understand cosmic phenomena that arises from the interaction between two or more individual universes, it seems that the multiverse, considered as a collective phenomena, may show features that were not contemplated so far. For instance, we have shown that, in terms of a new representation, the multiverse may be organized in a collective structure for which the energy levels available to single universes become quite different to those derived from the mere sum of universes. This shows that the multiverse is more than the sum of its parts, as it already happens in other branches of physics where interaction cannot be disregarded.

In what concerns the vacuum energy of a single universe, it is shown that it depends not only on its internal properties or even on an particular interaction between two or more universes, but it can depend on the structure of the multiverse as a whole. Thus, the logics of the multiverse becomes rather different from the one concerning a single universe, or even from the logics applied to a set of universes considered individually. We have shown that new energy levels for the multiverse have also been found even for the case where no interaction is explicitly present among the universes, which is a splendid example of the new physics that the quantum multiverse may conceal.

One weak link of our approach is that the discussed formalism is highly dependent on the chosen representation.  There seems to be then a degree of arbitrariness on the description of the quantum multiverse, which is actualy a general problem in quantum mechanics. Another difficulty is that we do not know how to properly interpret the found ``fermionic-bosonic" structure of the multiverse as discussed above. In any case, it is fairly interesting to see that the multiverse components can interact either as bosonic or as fermionic subsystems.

Despite all these shortcomings, our study opens the door to a wide range of new features of the multiverse, and the unravelled collective phenomena was shown to have a bearing on cosmological constant problem. Furthermore, the encountered collective properties of the multiverse do provide the possibility of lifting some of the objections raised in Ref. \cite{GEllis06} about the questionable predictive power of the multiverse hypothesis. Of course, other schemes could be considered to tackle the collective phenomena of coupled harmonic oscillators \cite{Kumar1980, Acebron1998, Pikovsky1999, Montbrio2004, Zanette2005, Kim2007, Lopez2007, Chimonidou2008, Almendral2010}. In this respect, it is of particular interest the study of  Refs. \cite{Audenaert2002, Plenio2004}, where the authors consider the dynamical evolution of the entanglement within a chain of harmonic oscillators, and find that quantum information processing mainly deals with the transfer quantum information between ``separated qbits". An approach along these lines could be tried in order to study the transfer of entanglement in the multiverse. This might be the focus of an ensuing future work.


\begin{thebibliography}{27}
\expandafter\ifx\csname natexlab\endcsname\relax\def\natexlab#1{#1}\fi
\expandafter\ifx\csname bibnamefont\endcsname\relax
  \def\bibnamefont#1{#1}\fi
\expandafter\ifx\csname bibfnamefont\endcsname\relax
  \def\bibfnamefont#1{#1}\fi
\expandafter\ifx\csname citenamefont\endcsname\relax
  \def\citenamefont#1{#1}\fi
\expandafter\ifx\csname url\endcsname\relax
  \def\url#1{\texttt{#1}}\fi
\expandafter\ifx\csname urlprefix\endcsname\relax\def\urlprefix{URL }\fi
\providecommand{\bibinfo}[2]{#2}
\providecommand{\eprint}[2][]{\url{#2}}

\bibitem[{\citenamefont{Everett}(1957)}]{Everett1957}
\bibinfo{author}{\bibfnamefont{H.}~\bibnamefont{Everett}},
  \bibinfo{journal}{Rev. Mod. Phys.} \textbf{\bibinfo{volume}{29}}
  (\bibinfo{year}{1957}).

\bibitem[{\citenamefont{Linde}(1986)}]{Linde1986}
\bibinfo{author}{\bibfnamefont{A.}~\bibnamefont{Linde}},
  \bibinfo{journal}{Phys. Lett. B} \textbf{\bibinfo{volume}{175}},
  \bibinfo{pages}{395} (\bibinfo{year}{1986}).

\bibitem[{\citenamefont{Bousso and Polchinski}(2000)}]{Bousso2000}
\bibinfo{author}{\bibfnamefont{R.}~\bibnamefont{Bousso}} \bibnamefont{and}
  \bibinfo{author}{\bibfnamefont{J.}~\bibnamefont{Polchinski}},
  \bibinfo{journal}{JHEP} \textbf{\bibinfo{volume}{0006}}, \bibinfo{pages}{006}
  (\bibinfo{year}{2000}), \eprint{arXiv:hep-th/0004134}.


\bibitem{Susskind1} L. Susskind, hep-th/0302219.


\bibitem{PolL} J. Polchinski, hep-th/0603249.


\bibitem[{\citenamefont{Holman et~al.}(2008)\citenamefont{Holman,
  Mersini-Houghton, and Takahashi}}]{Mersini2008}
\bibinfo{author}{\bibfnamefont{R.}~\bibnamefont{Holman}},
  \bibinfo{author}{\bibfnamefont{L.}~\bibnamefont{Mersini-Houghton}},
  \bibnamefont{and}
  \bibinfo{author}{\bibfnamefont{T.}~\bibnamefont{Takahashi}},
  \bibinfo{journal}{Phys. Rev. D} \textbf{\bibinfo{volume}{77}},
  \bibinfo{pages}{063510,063511} (\bibinfo{year}{2008}),
  \eprint{arXiv:hep-th/0611223, arXiv:hep-th/0612142}.

\bibitem{GEllis06} G.F.R. Ellis, astro-ph/0603266.


\bibitem{BoussoSusskind2011} R. Bousso and L. Susskind, arXiv:1105.3796 [hep-th].

\bibitem{Nomura2012} Y. Nomura, arXiv:1110.4630 [hep-th], arXiv:1205.2675 [hep-th].


\bibitem{OBHerdeiro2012} O.Bertolami and V. Herdeiro, in preparation. 


\bibitem[{\citenamefont{Bertolami}(2008)}]{Bertolami2008}
\bibinfo{author}{\bibfnamefont{O.}~\bibnamefont{Bertolami}},
  \bibinfo{journal}{Gen. Rel. Grav.} \textbf{\bibinfo{volume}{40}},
  \bibinfo{pages}{1891} (\bibinfo{year}{2008}), \eprint{arXiv:0705.2325}.

\bibitem[{\citenamefont{Linde}(1988)}]{Linde1988}
\bibinfo{author}{\bibfnamefont{A.}~\bibnamefont{Linde}},
  \bibinfo{journal}{Phys. Lett. B} \textbf{\bibinfo{volume}{200}},
  \bibinfo{pages}{272} (\bibinfo{year}{1988}).

\bibitem[{\citenamefont{Robles-P{\'e}rez
  et~al.}(2012)\citenamefont{Robles-P{\'e}rez, Alonso-Serrano, and
  Gonz{\'a}lez-D{\'\i}az}}]{RP2011}
\bibinfo{author}{\bibfnamefont{S.}~\bibnamefont{Robles-P{\'e}rez}},
  \bibinfo{author}{\bibfnamefont{A.}~\bibnamefont{Alonso-Serrano}},
  \bibnamefont{and} \bibinfo{author}{\bibfnamefont{P.~F.}
  \bibnamefont{Gonz{\'a}lez-D{\'\i}az}}, \bibinfo{journal}{Phys. Rev. D}
  \textbf{\bibinfo{volume}{85}}, \bibinfo{pages}{063511}
  (\bibinfo{year}{2012}), \eprint{arXiv:1111.3178}.

\bibitem[{\citenamefont{McGuigan}(1988)}]{McGuigan1988}
\bibinfo{author}{\bibfnamefont{M.}~\bibnamefont{McGuigan}},
  \bibinfo{journal}{Phys. Rev. D} \textbf{\bibinfo{volume}{38}},
  \bibinfo{pages}{3031} (\bibinfo{year}{1988}).

\bibitem[{\citenamefont{Rubakov}(1988)}]{Rubakov1988}
\bibinfo{author}{\bibfnamefont{V.~A.} \bibnamefont{Rubakov}},
  \bibinfo{journal}{Phys. Lett. B} \textbf{\bibinfo{volume}{214}},
  \bibinfo{pages}{503} (\bibinfo{year}{1988}).

\bibitem[{\citenamefont{Strominger}(1990)}]{Strominger1990}
\bibinfo{author}{\bibfnamefont{A.}~\bibnamefont{Strominger}}, in
  \emph{\bibinfo{booktitle}{Quantum Cosmology and Baby Universes}}, edited by
  \bibinfo{editor}{\bibfnamefont{S.}~\bibnamefont{Coleman}},
  \bibinfo{editor}{\bibfnamefont{J.~B.} \bibnamefont{Hartle}},
  \bibinfo{editor}{\bibfnamefont{T.}~\bibnamefont{Piran}}, \bibnamefont{and}
  \bibinfo{editor}{\bibfnamefont{S.}~\bibnamefont{Weinberg}}
  (\bibinfo{publisher}{World Scientific, London, UK}, \bibinfo{year}{1990}),
  vol.~\bibinfo{volume}{7}.

\bibitem[{\citenamefont{Macedo and Guedes}(2012)}]{Macedo2012}
\bibinfo{author}{\bibfnamefont{D.~X.} \bibnamefont{Macedo}} \bibnamefont{and}
  \bibinfo{author}{\bibfnamefont{I.}~\bibnamefont{Guedes}},
  \bibinfo{journal}{J. Math. Phys.} \textbf{\bibinfo{volume}{53}},
  \bibinfo{pages}{052101} (\bibinfo{year}{2012}).

\bibitem[{\citenamefont{Menouar et~al.}(2010)\citenamefont{Menouar, Maamache,
  and Choi}}]{Menouar2010}
\bibinfo{author}{\bibfnamefont{S.}~\bibnamefont{Menouar}},
  \bibinfo{author}{\bibfnamefont{M.}~\bibnamefont{Maamache}}, \bibnamefont{and}
  \bibinfo{author}{\bibfnamefont{J.~R.} \bibnamefont{Choi}}
  (\bibinfo{year}{2010}), \eprint{arXiv:1010.2015}.

\bibitem[{\citenamefont{Lievens et~al.}(2006)\citenamefont{Lievens, Stoilova,
  and Van~der Jeugt}}]{Lievens2006}
\bibinfo{author}{\bibfnamefont{S.}~\bibnamefont{Lievens}},
  \bibinfo{author}{\bibfnamefont{N.~I.} \bibnamefont{Stoilova}},
  \bibnamefont{and} \bibinfo{author}{\bibfnamefont{J.}~\bibnamefont{Van~der
  Jeugt}}, \bibinfo{journal}{J. Math. Phys.} \textbf{\bibinfo{volume}{47}},
  \bibinfo{pages}{113504} (\bibinfo{year}{2006}).

\bibitem[{\citenamefont{Lievens
  et~al.}(2008{\natexlab{a}})\citenamefont{Lievens, Stoilova, and Van~der
  Jeugt}}]{Lievens2008a}
\bibinfo{author}{\bibfnamefont{S.}~\bibnamefont{Lievens}},
  \bibinfo{author}{\bibfnamefont{N.~I.} \bibnamefont{Stoilova}},
  \bibnamefont{and} \bibinfo{author}{\bibfnamefont{J.}~\bibnamefont{Van~der
  Jeugt}}, \bibinfo{journal}{J. Math. Phys.} \textbf{\bibinfo{volume}{49}},
  \bibinfo{pages}{073502} (\bibinfo{year}{2008}{\natexlab{a}}),
  \eprint{arXiv:0709.0180}.

\bibitem[{\citenamefont{Lievens
  et~al.}(2008{\natexlab{b}})\citenamefont{Lievens, Stoilova, and Van~der
  Jeugt}}]{Lievens2008b}
\bibinfo{author}{\bibfnamefont{S.}~\bibnamefont{Lievens}},
  \bibinfo{author}{\bibfnamefont{N.~I.} \bibnamefont{Stoilova}},
  \bibnamefont{and} \bibinfo{author}{\bibfnamefont{J.}~\bibnamefont{Van~der
  Jeugt}}, \bibinfo{journal}{J. Phys. Conf. Series}
  \textbf{\bibinfo{volume}{128}}, \bibinfo{pages}{012028}
  (\bibinfo{year}{2008}{\natexlab{b}}).

\bibitem[{\citenamefont{Plenio et~al.}(2004)\citenamefont{Plenio, Hertley, and
  Eisert}}]{Plenio2004}
\bibinfo{author}{\bibfnamefont{M.~B.} \bibnamefont{Plenio}},
  \bibinfo{author}{\bibfnamefont{J.}~\bibnamefont{Hertley}}, \bibnamefont{and}
  \bibinfo{author}{\bibfnamefont{J.}~\bibnamefont{Eisert}},
  \bibinfo{journal}{New J. Phys.} \textbf{\bibinfo{volume}{6}},
  \bibinfo{pages}{36} (\bibinfo{year}{2004}).

\bibitem[{\citenamefont{Kumar and Mehta}(1980)}]{Kumar1980}
\bibinfo{author}{\bibfnamefont{S.}~\bibnamefont{Kumar}} \bibnamefont{and}
  \bibinfo{author}{\bibfnamefont{C.~L.} \bibnamefont{Mehta}},
  \bibinfo{journal}{J. Math. Phys.} \textbf{\bibinfo{volume}{21}},
  \bibinfo{pages}{2628} (\bibinfo{year}{1980}).

\bibitem[{\citenamefont{Acebr{\'o}n and Bonilla}(1998)}]{Acebron1998}
\bibinfo{author}{\bibfnamefont{J.~A.} \bibnamefont{Acebr{\'o}n}}
  \bibnamefont{and} \bibinfo{author}{\bibfnamefont{L.~L.}
  \bibnamefont{Bonilla}}, \bibinfo{journal}{Physica D}
  \textbf{\bibinfo{volume}{114}}, \bibinfo{pages}{296} (\bibinfo{year}{1998}).

\bibitem[{\citenamefont{Pikovsky and Ruffo}(1999)}]{Pikovsky1999}
\bibinfo{author}{\bibfnamefont{A.}~\bibnamefont{Pikovsky}} \bibnamefont{and}
  \bibinfo{author}{\bibfnamefont{S.}~\bibnamefont{Ruffo}},
  \bibinfo{journal}{Phys. Rev. E} \textbf{\bibinfo{volume}{59}},
  \bibinfo{pages}{1633} (\bibinfo{year}{1999}).

\bibitem[{\citenamefont{Montbri{\'o} et~al.}(2004)\citenamefont{Montbri{\'o},
  Kurths, and Blasius}}]{Montbrio2004}
\bibinfo{author}{\bibfnamefont{E.}~\bibnamefont{Montbri{\'o}}},
  \bibinfo{author}{\bibfnamefont{J.}~\bibnamefont{Kurths}}, \bibnamefont{and}
  \bibinfo{author}{\bibfnamefont{B.}~\bibnamefont{Blasius}},
  \bibinfo{journal}{Phys. Rev. E} \textbf{\bibinfo{volume}{70}},
  \bibinfo{pages}{056125} (\bibinfo{year}{2004}).

\bibitem[{\citenamefont{Zanette}(2005)}]{Zanette2005}
\bibinfo{author}{\bibfnamefont{D.~H.} \bibnamefont{Zanette}},
  \bibinfo{journal}{Eur. Phys. J. B} \textbf{\bibinfo{volume}{43}},
  \bibinfo{pages}{97} (\bibinfo{year}{2005}).

\bibitem[{\citenamefont{Kim and Noz}(2007)}]{Kim2007}
\bibinfo{author}{\bibfnamefont{S.~P.} \bibnamefont{Kim}} \bibnamefont{and}
  \bibinfo{author}{\bibfnamefont{M.~E.} \bibnamefont{Noz}},
  \bibinfo{journal}{J. Opt. B Quant. Semiclass. Opt}
  \textbf{\bibinfo{volume}{7}}, \bibinfo{pages}{S458} (\bibinfo{year}{2007}).

\bibitem[{\citenamefont{L{\'o}pez et~al.}(2007)\citenamefont{L{\'o}pez,
  L{\'o}pez, and L{\'o}pez}}]{Lopez2007}
\bibinfo{author}{\bibfnamefont{G.~V.} \bibnamefont{L{\'o}pez}},
  \bibinfo{author}{\bibfnamefont{P.}~\bibnamefont{L{\'o}pez}},
  \bibnamefont{and} \bibinfo{author}{\bibfnamefont{X.~E.}
  \bibnamefont{L{\'o}pez}}, \bibinfo{journal}{Int. J. Theor. Phys.}
  \textbf{\bibinfo{volume}{46}}, \bibinfo{pages}{1100} (\bibinfo{year}{2007}).

\bibitem[{\citenamefont{Chimonidou and Sudarshan}(2008)}]{Chimonidou2008}
\bibinfo{author}{\bibfnamefont{A.}~\bibnamefont{Chimonidou}} \bibnamefont{and}
  \bibinfo{author}{\bibfnamefont{E.~C.~G.} \bibnamefont{Sudarshan}},
  \bibinfo{journal}{Phys. Rev. A} \textbf{\bibinfo{volume}{77}},
  \bibinfo{pages}{032121} (\bibinfo{year}{2008}).

\bibitem[{\citenamefont{Almendral et~al.}(2010)}]{Almendral2010}
\bibinfo{author}{\bibfnamefont{J.~A.} \bibnamefont{Almendral}}
  \bibnamefont{et~al.}, \bibinfo{journal}{Int. J. Bifurc. Chaos}
  \textbf{\bibinfo{volume}{20}}, \bibinfo{pages}{753} (\bibinfo{year}{2010}).

\bibitem[{\citenamefont{Audenaert et~al.}(2002)\citenamefont{Audenaert, Eisert,
  and Plenio}}]{Audenaert2002}
\bibinfo{author}{\bibfnamefont{K.}~\bibnamefont{Audenaert}},
  \bibinfo{author}{\bibfnamefont{J.}~\bibnamefont{Eisert}}, \bibnamefont{and}
  \bibinfo{author}{\bibfnamefont{M.~B.} \bibnamefont{Plenio}},
  \bibinfo{journal}{Phys. Rev. A} \textbf{\bibinfo{volume}{66}},
  \bibinfo{pages}{042327} (\bibinfo{year}{2002}).

\end{thebibliography}

\end{document}